\documentclass[a4paper]{article}

\usepackage{INTERSPEECH2021}
 \usepackage{mathrsfs}
 \usepackage[bookmarks=false]{hyperref}
 \usepackage{xcolor}
 \usepackage{booktabs}
 \usepackage{array}
 \usepackage{ulem} 
\usepackage{cleveref}
\usepackage{cite}
\title{Applying the Information Bottleneck Principle to Prosodic Representation Learning}
\name{Guangyan Zhang$^1$, Ying Qin$^2$, Daxin Tan$^1$, Tan Lee$^1$}
\address{
  $^1$ Department of Electronic Engineering, The Chinese University of Hong Kong\\
  $^2$ Institute of Information Science, Beijing Jiaotong University, Beijing 100044, China}
\email{\{gyzhang, daxintan\}@link.cuhk.edu.hk, yingqin@bjtu.edu.cn, tanlee@cuhk.edu.hk}

\begin{document}

\maketitle
\vspace{-1em}
\begin{abstract}
This paper describes a novel design of a neural network-based speech generation model for learning prosodic representation. The problem of representation learning is formulated according to the information bottleneck (IB) principle. A modified VQ-VAE quantized layer is incorporated in the speech generation model to control the IB capacity and adjust the balance between reconstruction power and disentangle capability of the learned representation. The proposed model is able to learn word-level prosodic representations from speech data. With an optimized IB capacity, the learned representations not only are adequate to reconstruct the original speech but also can
be used to transfer the prosody onto different textual content. Extensive results of the objective and subjective evaluation are presented to demonstrate the effect of IB capacity control, the effectiveness, and potential usage of the learned prosodic representation in controllable neural speech generation.

\end{abstract}

\vspace{-1em}
\section{Introduction}
Prosody is an essential component embedded in human speech. Carrying a full array of linguistic and paralinguistic functions, prosody manifests tone, stress, intonation, and rhythm of speech and contributes to naturalness, style, attitude, and emotion of speech \cite{taylor2009text,wagner2010experimental}. Conventionally speech prosody has been studied through the analysis of pitch, intensity, and duration features, focusing on one or more specific functions of prosody. This approach does not provide a holistic representation of prosody for speech generation purposes. In recent years, neural network-based speech synthesis systems \cite{wang2017tacotron,shen2018natural,li2019neural} show superior performance in terms of speech quality and naturalness and offer an effective approach to learning speech representations. The problem of prosodic representation learning has attracted particular attention in view of its potential use in prosody transfer and style control \cite{skerry2018towards,klimkov2019fine,lee2019robust}, prosody control \cite{zhang2020learning,hodariusing}.

Various approaches to unsupervised learning of prosodic representation have been proposed \cite{skerry2018towards,klimkov2019fine,lee2019robust,zhang2020learning,sun2020fully,tan2020fine,wang2019vector,kenter2019chive}. In these studies, a neural network model is used to disentangle contributing factors of input speech and subsequently perform speech reconstruction from the disentangled factors. The prosodic representation is obtained as one of the learned factors, parallel with non-prosodic factors that correspond to content, speaker, channel, etc. In
\cite{tan2020fine, zhang2021estimating, hsu2019disentangling,karlapati2020copycat},  adversarial learning was applied to address the problem that the learned prosodic representation might contain substantial information related to non-prosodic factors. The use of an adversarial classifier requires the availability of the labels for one of the disentangled non-prosodic factors. The design of the adversarial classifier is specific to only one non-prosodic factor and can not be applied to other non-prosodic factors. Furthermore, the non-prosodic factors(e.g., speaker) might be related to prosody  \cite{zhang2021estimating}, while disentangling with an adversarial classifier might also result in low prosody information in the prosodic representation.  

In the present study, prosodic representations learning is tackled from the perspective of information bottleneck (IB), by which a good representation is determined with the trade-off between its predictive/reconstructive power and compact representation \cite{tishby2015deep}. We propose to define a good prosodic representation in three different aspects. First, it should have a good capability of capturing prosody-related information. It can reconstruct the reference speech conditioned on the other necessary factors (good reconstructive power). Second, the representation is expected to include as little as possible information about non-prosodic factors (compact representation). Third, the predicted prosodic representation should contribute to generating natural speech. The prosodic representations extracted from reference speech could be used to train a prosody predictor, which predicts the prosodic representation from the text. Specific speech generation applications (e.g., TTS) require appropriate and expressive prosody predicted from the text \cite{hodariusing}. The predicted prosodic representation is expected to help improve the naturalness or expressiveness of the generated speech.

The contributions of this study are as follows: (1) the prosodic representation learning problem is formulated based on the IB principle; (2) a prosodic representation learning system with controllable IB capacity is developed; (3) subjective and objective evaluation results show that learned prosodic representations have good reconstructive power and can concisely capture prosody-related information by choosing appropriate IB capacity; (4) a machine-translation-based prosody predictor is proposed to realize text-to-prosody prediction. The predictor can generate prosodic representations for improving the naturalness of synthesized speech.

\vspace{-1em}
\section{Problem Formulation}
\label{sec:formlula}
The IB principle aims to learn robust representation from data for a specific task, e.g., classification, auto-encoding \cite{alemi2016deep}.
Let $Z$ denote the representation to be learned from data $X$, and $Y$ denote the task target\footnote{For auto-encoding, the task target $Y$ is the same as $X$.}. The goal of learning is to maximize the mutual information between $Z$ and $Y$, and meanwhile, to discard irrelevant information about $Y$ that might be present in the data $X$.
Let $I(Z;Y )$ and $I(X;Z )$ be the mutual information between $Z$ and $Y$ and that between $X$ and $Z$ respectively. That is, the following objective function is to be maximized \cite{tishby2015deep,tishby2000information,chechik2005information},
\begin{equation}
  I(Z;Y ) - \beta I(X; Z ),\footnote{In this paper, $X,Y,Z,F_i$ are random variables and $x,z,t$ are instances of random variables.} 
\end{equation}
where $\beta$ is a Lagrange multiplier.

Recent studies \cite{skerry2018towards,klimkov2019fine,lee2019robust,zhang2020learning,sun2020fully,tan2020fine,wang2019vector,kenter2019chive} share similar strategies for learning prosodic representation $Z$ from speech data $X$. The learning model encodes input $X$ into prosodic representation $Z$, and reconstructs $X$ from $Z$ in conjunction with the representations of non-prosodic factors $F_i$, e.g., speaker, channel, speech content. Following the information bottleneck framework, prosodic representation learning is achieved by maximizing
\vspace{-0.4em}
\begin{equation}
  I(Z;X|F_1,F_2,\cdots,F_i,\cdots,F_N) - \beta I(X; Z ).
  \label{eq_ib2}
\end{equation} 
\vspace{-0.2em}
With the first term in \autoref{eq_ib2}, $Z$ is expected to have the capability of reconstructing $X$, given $F_i$.  The second term in \autoref{eq_ib2} is introduced as a constraint to prevent $Z$ from being a direct copy of $X$. The present study is carried out with a single-speaker corpus, and it is assumed that content is the only non-prosodic factor needed.

The main challenge of applying the IB principle is on the computation of mutual information. Variational inference is regarded as a practical way to approximate the calculation \cite{alemi2016deep}. In this study, variational inference is adopted to establish a lower bound on the IB objective function  \cite{alemi2016deep,alemi2018fixing} in \autoref{eq_ib2}. Let $x$, $z$ and $t$ be the speech, prosodic representation, and content factor, respectively. $q_\phi(z|x)$ is the variational approximation of the true posterior $p_\theta(z|x)$, which is implemented by a reference encoder as described in \autoref{Section:ref_encoder}. The conditional probability $p_\theta(x|z,t)$ is modeled by a conditional decoder which aims to reconstruct speech (\autoref{Section:con_decoder}). $p(z)$ is the variational approximation to the marginal distribution of $z$.  The approximation of \autoref{eq_ib2} gives the objective function,
\begin{equation}
  E_{q_\phi(z|x)}[\log p_\theta(x|z,t)] - \beta D_{KL}(q_\phi(z|x)||p(z)).
  \label{eq:beta_vae}
\end{equation} 
The first term in \autoref{eq:beta_vae} represents the reconstruction task. In the second term of \autoref{eq:beta_vae}, $D_{KL}(q_\phi(z|x)||p(z))$ indicates the KL divergence between $q_\phi(z|x)$ and $p(z)$. The  KL divergence $D_{KL}(q_\phi(z|x)||p(z))$, known as the IB capacity, is the upper bound of the mutual information $I(X; Z)$ in \autoref{eq_ib2} \cite{higgins2017learning,burgess2018understanding}.  Control of information transmitted from speech to prosodic representation can be achieved by adjusting the IB capacity. With limited IB capacity, the prosodic representation is expected to contain mostly prosody information. As the IB capacity increases, information of other factors would be absorbed in the prosodic representation.

\begin{figure*}[t]
  \centering
  \includegraphics[width=\textwidth]{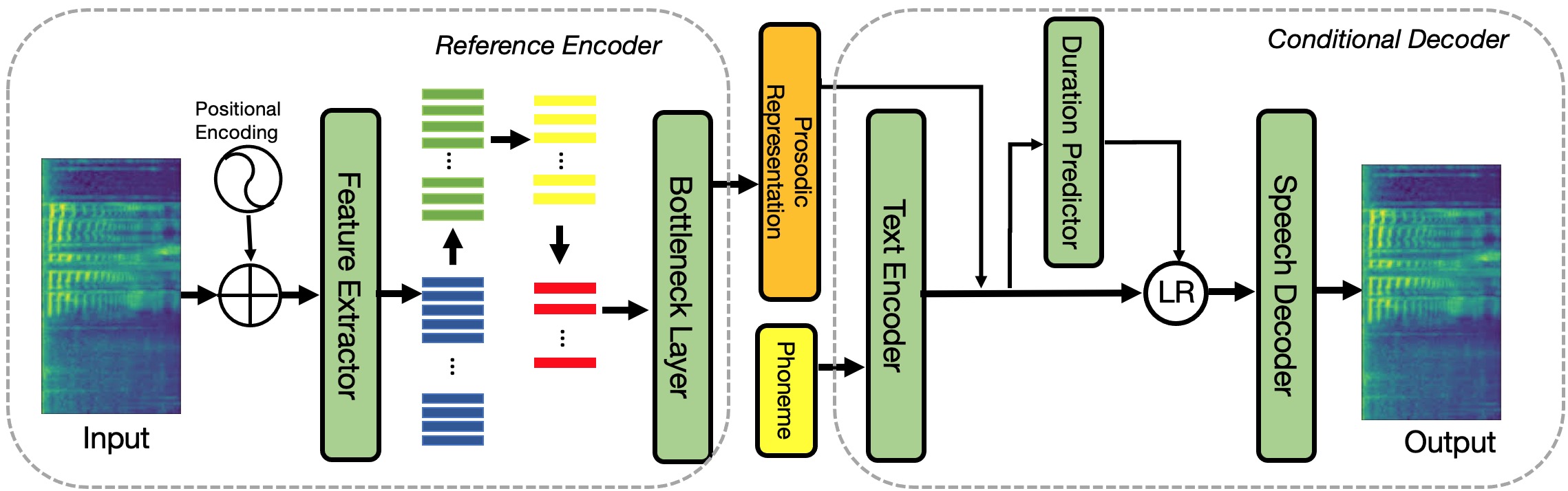}
  \caption{The structure of prosodic representation learning system. The vectors in blue, green, yellow, and red colors represent frame, phoneme, syllable and word-level acoustic features respectively. }
  \label{Figure:model_arch}
\end{figure*}

\vspace{-0.5em}
\section{Model Design}
\label{sec:model_design}
The proposed system consists of a reference encoder and a conditional decoder, as shown in \autoref{Figure:model_arch}. In the present study, the prosodic representation is learned at the word level. The reference encoder extracts a sequence of word-level prosodic representations with controlled IB capacity from an input mel-spectrogram. The conditional decoder reconstructs the mel-spectrogram from the prosodic representation.  
\vspace{-0.3em}
\subsection{Reference Encoder}
\label{Section:ref_encoder}
The reference encoder is made up of a feature extractor and a bottleneck layer. The feature extractor takes mel-spectrogram as input and generates word-level acoustic features. These features are processed by the bottleneck layer and encoded into a sequence of prosodic representations.
\vspace{-0.2em}
\subsubsection{Feature Extractor}
 The feature extractor contains four Feed-Forward Transformer (FFT) \cite{ren2019fastspeech} blocks. Each FFT block consists of a self-attention \cite{vaswani2017attention} and a 1D convolutional network. The input mel-spectrogram is first added with sinusoid positional encoding, then passed through the feature extractor. The feature extractor outputs frame-level acoustic features.   The frame-level features are aggregated to phoneme-level features by average pooling. In the same manner, the syllable and word-level acoustic features are obtained.  
\vspace{-0.2em}
\subsubsection{Bottleneck Layer}
\label{bottleneck_layer}
The bottleneck layer takes word-level acoustic features as input and outputs prosodic representations. It aims to remove non-prosodic information (e.g., speaker, content, and channel). In \cite{burgess2018understanding}, the bottleneck layer was implemented with the standard Variational AutoEncoder (VAE) \cite{kingma2014auto}. However, this leads a notorious issue when the bottleneck layer is trained on sequential data (e.g., speech or text): the output representation forgets all information regarding the data and has the identical distribution as the uninformative marginal prior,
yielding the so-called KL vanishing problem \cite{fu2019cyclical}. In this paper, the bottleneck layer is implemented by a modified Vector Quantized VAE (VQ-VAE) quantized layer \cite{van2017neural}. The VQ-VAE quantized layer has a deterministic latent representation and does not have the problem of KL vanishing. VQ-VAE realizes a bottleneck that quantizes input feature, directly limiting the amount of information embedded in learned representation. It learns a dictionary (codebook) $\mathbf{E} \in \mathbb{R}^{K \times D}$, where $K$ is the dictionary size, and $D$ is the dimension of each code $\mathbf{e}_i$. The continuous feature $\mathbf{x}$ can be encoded as the discrete code $z$ using the nearest neighbor lookup. The variational posterior  $q_\phi(z|\mathbf{x})$ is deterministic and defined as, 
\vspace{-0.5em}
 \begin{equation}
\label{eq6}
q_\phi(z=k|\mathbf{x}) = \left\{
\begin{aligned}
1 &, & \text{for } k=\arg \min_i||\mathbf{x}-\mathbf{e}_i||_2, \\
0 & , & \text{otherwise}.
\end{aligned}
\right.
\end{equation}
\vspace{-0.2em}
The marginal distribution $p(z)$ based on variational approximation is assumed to be a simple uniform distribution. The KL divergence between $q_\phi(z|\mathbf{x})$ and $p(z)$ is equal to $\log K$.

The IB capacity of the VQ-VAE quantized layer (i.e., $\log K$) can be controlled by changing the dictionary size $K$. The larger the size $K$, the higher the IB capacity. However, increasing the dictionary size would cause the out-of-memory problem.  The coding strategy in \cite{baevski2019vq} was used to mitigate the mode collapse problem in VQ-VAE. It can increase the bottleneck layer IB capacity without taking up extra memory,but the IB capacity cannot be controlled and the coding strategy has not been used for learning prosodic representations. In this study, a novel prosodic representation learning system with controllable IB capacity is proposed. 

The word-level prosodic representation can be encoded with multiple entries in the dictionary $\mathbf{E}$. The acoustic feature $\mathbf{x} \in \mathbb{R}^{D}$ is first divided equally into $G$ groups and arranged as a matrix $ \mathbf{X}'=[\mathbf{x}_1,\mathbf{x}_2,\cdots,\mathbf{x}_j,\cdots \mathbf{x}_G]\in \mathbb{R}^{(D/G) \times G}$. Each column of $\mathbf{X}'$ can be encoded by an integer index independently as the prosodic representation  $\mathbf{z}'=[z_1,z_2,\cdots,z_j,\cdots, z_G] \in \mathbb{Z}^{1 \times G}$. The variational posterior distribution $q_\phi(\mathbf{z}'|\mathbf{X}')$ and marginal distribution $p(\mathbf{z}')$ can be represented as,

\vspace{-1.2em}
\begin{equation}
\begin{aligned}
q_\phi(\mathbf{z}'|\mathbf{X}') &= q_\phi(z_1,z_2,\cdots,z_j,\cdots,z_G|\mathbf{X}') \\
& = \prod_{j=1}^G q_\phi(z_j=k|\mathbf{x}_j),
 \end{aligned}
\end{equation}
\vspace{-1.5em}
\begin{equation}
p(\mathbf{z}') = p(z_1,z_2,\cdots,z_j,\cdots,z_G) = \prod_{j=1}^G p(z_j).
\end{equation}
Then, the $D_{KL}(q_\phi(\mathbf{z}'|\mathbf{X}')||p(\mathbf{z}'))$ can be derived as,
\begin{equation}
    \begin{aligned}
    D_{KL}(q_\phi(\mathbf{z}'|\mathbf{X}')||p(\mathbf{z}')) &= \sum_{\mathbf{z}'} q_\phi(\mathbf{z}'|\mathbf{X}')\log\frac{q_\phi(\mathbf{z}'|\mathbf{X}')}{p(\mathbf{z}')} \\
    &=\sum_{g=1}^G \sum_{z_g=1}^K q_\phi(z_g|\mathbf{x}_g) \log\frac{q_\phi(z_g|\mathbf{x}_g)}{p(z_g)} \\
   & = G \log K.
    \end{aligned}
\end{equation}
In this paper, the group number $G$ is set as 2.

\subsection{Conditional Decoder}
\label{Section:con_decoder}
The design of the conditional decoder follows the Fastspeech2 system with some adjustment \cite{ren2020fastspeech}. It comprises a text encoder and a speech decoder.  The phone sequence is passed through the text encoder to obtain phone-level features. The word-level prosodic representations are up-sampled to match the granularity of phone level and concatenated to the phone-level features. The Length Regulator (LR) module further up-samples all phone-level features to frame-level features, which are presented to the speech decoder to generate mel-spectrogram. The phone-to-frame up-sampling operation requires phone duration information provided by forced alignment in the training stage. At the inference stage, the duration is predicted with a duration predictor.
\vspace{-1.5em}

\section{Performance Evaluation}
\label{evaluation}
\subsection{Experimental Setup}

A series of experiments are carried out with the part of Blizzard 2013 English dataset. It contains approximately $57$ hours of recordings covering a few non-fiction audiobooks in an expressive reading style. Phone-level time alignment and word-to-syllable conversion were performed using the Festival software toolkits\cite{black1998festival}. The Parallel WaveGAN vocoder \cite{yamamoto2020parallel} is used to generate speech waveform from mel-spectrograms.

Different settings of IB capacities are experimented. The capacity is changed by varying the dictionary size $K$ as described in \autoref{bottleneck_layer}. The dictionary sizes and corresponding IB capacities are shown in \autoref{ib_table}.

\begin{table}[htbp]
        \caption{Different dictionary sizes and corresponding IB capacities set for bottleneck layer.  }
        \vspace{-1.0em}
    \centering
    \begin{tabular}{m{2.2cm}m{0.15cm}m{0.38cm}m{0.38cm}m{0.38cm}m{0.38cm}m{0.38cm}m{0.38cm}m{0.38cm}}
         \toprule 
         dictionary size & 0 & 2 & 4 &8 & 16 & 32 & 64 \\
         \midrule
         IB capacity(nats)& $0$ & $1.39$ & $2.77$ & $4.16$ & $5.54$ & $6.93$ & $8.31$ \\
         \bottomrule
    \end{tabular}
    \label{ib_table}
    \vspace{-1.0em}
\end{table}

\vspace{-0.5em}
\subsection{Reconstruction of Speech}
\vspace{-0.5em}
\label{recon_speech}
In this task, the reference speech is reconstructed by the conditional decoder using the learned prosodic representations and the phone sequence from reference speech. The reconstruction performance reflects whether the learned prosodic representation carries enough prosody information. Four metrics that are related to prosody in the acoustic aspect, including  Voicing  Decision  Error  (VDE),  Gross Pitch Error (GPE),  F0 Frame Error (FFE), and Mel Cepstral distortion ($\text{MCD}_{22}$) are used to evaluate the reconstruction performance \cite{skerry2018towards}. The results are shown as in \autoref{recon_table}, where smaller values mean better performance.  As the IB capacity is increased from 0nats to 8.31nats, the reconstruction performance improves. The values of all four metrics drop rapidly as the capacity starts to increase. This suggests that the information passed to the prosodic representation is most useful for reconstruction and highly related to speech prosody. As the capacity continues to increase, the degree of performance improvement becomes less significant. For the range of 6.93 nats to 8.31nats, the GPE, FFE, $\text{MCD}_{22}$ fall slightly by $0.26\%$, $0.51\%$, and $0.09$, respectively. It suggests that the information passed to the prosodic representation has a low correlation with prosody as the IB capacity is large. Readers are recommended to listen to demo examples \footnote{https://patrick-g-zhang.github.io/pt/}.

\begin{table}[htbp]
        \caption{Objective evaluation on speech reconstruction. }
        \vspace{-1.0em}
    \centering
    \begin{tabular}{m{2.2cm}m{0.8cm}m{0.8cm}m{0.7cm}m{0.7cm}}
         \toprule
         IB capacity(nats) & VDE($\%$) & GPE($\%$) & FFE($\%$) &$\text{MCD}_{22}$ \\
         \midrule
         0 & $15,14$ & $39.10$ & $37.39$ & $6.74$\\
         1.39 & $11.48$ & $27.70$ & $27.39$ & $6.13$  \\
         2.77 & $10.83$ & $13.21$ & $18.30$ & $5.69$ \\
         4.16 & $10.21$ & $9.05$  & $15.38$ & $5.57$ \\
         5.54 & $\mathbf{9.37}$ & $7.56$ & $13.72$ & $5.43$ \\
         6.93 & $9.89$ & $6.40$ & $13.47$ & $5.17$ \\
         8.31 & $9.42$ & $\mathbf{6.14}$ & $\mathbf{12.96}$ & $\mathbf{5.08}$ \\
         \bottomrule
    \end{tabular}
    \label{recon_table}
\end{table}
\vspace{-1em}
\subsection{Mutual Information between Content and Prosodic Representation}
\begin{figure}[t]
  \centering
  \includegraphics[width=0.7\linewidth]{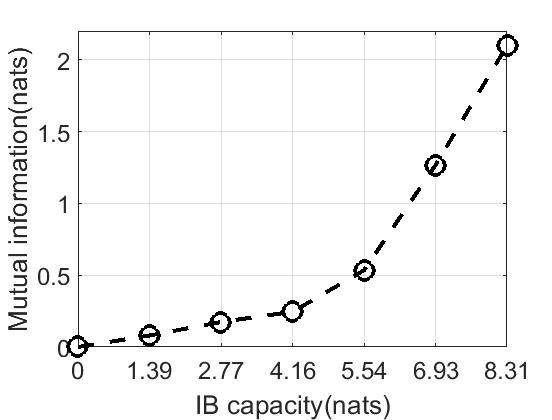}
  \caption{Mutual information between content and prosodic representation.}
  \label{fig:mi_text}
\end{figure}

In \Cref{recon_speech}, it is noted that the IB capacity should be properly controlled to reach a balance between reconstruction performance and the efficacy of prosodic representation.  Hence the mutual information between content and prosodic representation is evaluated.  The phone embeddings in each word are aggregated to form a continuous-valued vector.  The mutual information between this word-level vector and the respective prosodic representation is estimated by MINE \cite{belghazi2018mutual}. The results of the evaluation are shown as in \autoref{fig:mi_text}. It can be seen that the mutual information increases mildly as the IB capacity is in the range of 0 to 5.54 nats, meaning that the prosodic representation does not entangle with much content information. The mutual information increases sharply as the IB capacity exceeds 5.54 nats. In this case, content information is leaked into the learned prosodic representation.

\vspace{-1.0em}
\subsection{Cross-text Prosody Transfer}

The efficacy of learned prosodic representation can be evaluated in cross-text prosody transfer, where speech is generated using the prosodic representation and a given phone sequence which is different from reference speech. If the prosodic representation is highly entangled with the content factor, the content of generated speech content would be corrupted. Subjective evaluation was carried out by a listening test implemented via the Amazon Mechanical Turk platform. $20$ native English speakers participated in the listening test. There were two tasks in the test. One of them aimed to evaluate the effectiveness of prosody transfer. The listener was asked to give a 5-point  preference score on the prosody similarity between the reference speech and the generated speech, ranging from ``completely different'' (1) to ``completely same'' (5). The other task was designed to evaluate the clearness of generated speech content. Each listener was required to give a score that rates the clearness of generated speech content. The score ranges from ``completely unclear'' (1) to ``completely clear'' (5) with five levels. 

Results of the subjective evaluation are shown as in \autoref{cross_table}. In terms of prosody similarity, the score grows initially and then saturates as the IB capacity increases. The clearness of content gradually declines as the IB capacity increases and degrades significantly when the IB capacity increases from 5.54 nats to 6.93  nats. By examining the test samples of generated speech, it is found that the deterioration of clearness is mainly caused by slurred words. At the IB capacity of 5.54 nats (dictionary size of 16), the learned prosodic representation achieves the best performance in subjective evaluation. This represents a balance between prosodic and non-prosodic factors embedded in the prosodic representation.

\begin{table}[htbp]
        \caption{Subjective evaluation scores of cross-text prosody transfer. }
        \vspace{-1.0em}
    \centering
    \begin{tabular}{lcccc}
         \toprule  
         IB capacity(nats) & prosody similarity & content clearness \\
         \midrule
         0 & $1.87 \pm 0.73$ & $\mathbf{3.61 \pm 0.88}$ \\
         1.39 & $2.33 \pm 0.84$ & $3.53 \pm 0.94$ \\
         2.77 & $2.96 \pm 0.82$ & $3.50 \pm 0.89$ \\
         4.16 & $3.12 \pm 0.86 $ & $3.49 \pm 0.90$ \\
         5.54 & $\mathbf{3.25 \pm 0.84}$ & $3.49 \pm 0.94$ \\
         6.93 & $3.24 \pm 0.86$ & $3.23 \pm 0.96$ \\
         8.31 & $\mathbf{3.25 \pm 0.89}$ & $3.17 \pm 0.90$ \\
         \bottomrule
    \end{tabular}
    \label{cross_table}
\end{table}

\vspace{-1.0em}
\subsection{Predicting Prosodic Representation from Text }
In a trained system for prosodic representation learning, prosodic representation can be extracted by the reference encoder. A prosody predictor is proposed to predict the prosodic representation given a text sequence. The input of the prosody predictor is given by the contextual embedding extracted from the BERT \cite{devlin2019bert,jawahar2019does}, which incorporates rich syntactic and semantic information. The output of the prosody predictor is the prosodic representation. The prosody predictor model follows the transformer-based machine translation system \cite{ott2019fairseq} with both text and prosodic representation being discrete. The prosody predictor is trained with prosodic representations extracted with different IB capacities settings. The predicted prosodic representations are then passed to the conditional decoder to generate speech. A MOS test was carried out to evaluate the naturalness of generated speech. The score ranges from ``completely unnatural" (1) to ``completely natural" (5) with five levels. The MOS result is shown as in \autoref{mos_test}. Compared to the case with no input to the conditional decoder (i.e., IB capacity of 0), the predicted prosodic representations with IB capacity greater than 0 could improve the naturalness of generated speech. The predicted prosodic representations with an IB capacity of 5.54 nats (dictionary size of 16) achieve the highest score among all settings. It suggests that with appropriate control of IB capacity, the predicted prosodic representation could help a TTS system improve the naturalness of output speech.
\begin{table}[htbp]
        \caption{MOS test result.  }
        \vspace{-1.0em}
    \centering
    \begin{tabular}{m{2.2cm}m{0.35cm}m{0.35cm}m{0.35cm}m{0.35cm}m{0.35cm}m{0.35cm}m{0.35cm}m{0.35cm}|}
         \toprule
         IB capacity(nats)& $0$ & $1.39$ & $2.77$ & $4.16$ & $5.54$ & $6.93$ & $8.31$ \\
          \midrule 
         MOS & 3.75 & 3.79 & 3.80 &3.87 & \textbf{3.93} & 3.92 & 3.85 \\
         \bottomrule
    \end{tabular}
    \label{mos_test}
\end{table}

\vspace{-1.5em}
\section{Conclusions}

The information bottleneck (IB) principle is applied to learning word-level prosodic representation from speech data. Through proper control of the IB capacity, the learned prosodic representations, on the one hand, are adequate for speech reconstruction and, on the other hand, compactly capture prosody-related information in speech. The predicted prosodic representation from the prosody predictor has shown the potential to improve the naturalness of speech for existing TTS systems. In our future work, speaker or channel factors will be investigated in the multi-speaker and multi-channel scenarios.


\section{Acknowledgement}

This research is partially supported by a Tier 3 funding from ITSP (Ref: ITS/309/18) of the Hong Kong SAR Government, and a Knowledge Transfer Project Fund (Ref: KPF20QEP26) from the Chinese University of Hong Kong.

\bibliographystyle{IEEEtran}

\bibliography{mybib}

\begin{thebibliography}{10}
\providecommand{\url}[1]{#1}
\csname url@samestyle\endcsname
\providecommand{\newblock}{\relax}
\providecommand{\bibinfo}[2]{#2}
\providecommand{\BIBentrySTDinterwordspacing}{\spaceskip=0pt\relax}
\providecommand{\BIBentryALTinterwordstretchfactor}{4}
\providecommand{\BIBentryALTinterwordspacing}{\spaceskip=\fontdimen2\font plus
\BIBentryALTinterwordstretchfactor\fontdimen3\font minus
  \fontdimen4\font\relax}
\providecommand{\BIBforeignlanguage}[2]{{%
\expandafter\ifx\csname l@#1\endcsname\relax
\typeout{** WARNING: IEEEtran.bst: No hyphenation pattern has been}%
\typeout{** loaded for the language `#1'. Using the pattern for}%
\typeout{** the default language instead.}%
\else
\language=\csname l@#1\endcsname
\fi
#2}}
\providecommand{\BIBdecl}{\relax}
\BIBdecl

\bibitem{taylor2009text}
P.~A. Taylor, \emph{Text-to-speech synthesis}.\hskip 1em plus 0.5em minus
  0.4em\relax Cambridge University Press, 2009.

\bibitem{wagner2010experimental}
M.~Wagner and D.~G. Watson, ``Experimental and theoretical advances in prosody:
  A review,'' \emph{Language and cognitive processes}, vol.~25, no. 7-9, pp.
  905--945, 2010.

\bibitem{wang2017tacotron}
Y.~Wang, R.~Skerry-Ryan, D.~Stanton, Y.~Wu, R.~J. Weiss, N.~Jaitly, Z.~Yang,
  Y.~Xiao, Z.~Chen, S.~Bengio \emph{et~al.}, ``Tacotron: Towards end-to-end
  speech synthesis,'' in \emph{Proc. Interspeech}, Aug. 2017, pp. 4006--4010.

\bibitem{shen2018natural}
J.~{Shen}, R.~{Pang}, R.~J. {Weiss}, M.~{Schuster}, N.~{Jaitly}, Z.~{Yang},
  Z.~{Chen}, Y.~{Zhang}, Y.~{Wang}, R.~{Skerrv-Ryan}, R.~A. {Saurous},
  Y.~{Agiomvrgiannakis}, and Y.~{Wu}, ``Natural tts synthesis by conditioning
  wavenet on mel spectrogram predictions,'' in \emph{Proc. ICASSP}, 2018, pp.
  4779--4783.

\bibitem{li2019neural}
N.~Li, S.~Liu, Y.~Liu, S.~Zhao, and M.~Liu, ``Neural speech synthesis with
  transformer network,'' in \emph{Proc. AAAI}, vol.~33, no.~01, 2019, pp.
  6706--6713.

\bibitem{skerry2018towards}
R.~Skerry-Ryan, E.~Battenberg, Y.~Xiao, Y.~Wang, D.~Stanton, J.~Shor, R.~J.
  Weiss, R.~Clark, and R.~A. Saurous, ``Towards end-to-end prosody transfer for
  expressive speech synthesis with tacotron,'' in \emph{Proc. ICML}, 2018, pp.
  4700--4709.

\bibitem{klimkov2019fine}
V.~Klimkov, S.~Ronanki, J.~Rohnke, and T.~Drugman, ``Fine-grained robust
  prosody transfer for single-speaker neural text-to-speech,'' in \emph{Proc.
  Interspeech}, 2019, pp. 4440--4444.

\bibitem{lee2019robust}
Y.~Lee and T.~Kim, ``Robust and fine-grained prosody control of end-to-end
  speech synthesis,'' in \emph{Proc. ICASSP}, 2019, pp. 5911--5915.

\bibitem{zhang2020learning}
G.~Zhang, Y.~Qin, and T.~Lee, ``Learning syllable-level discrete prosodic
  representation for expressive speech generation,'' in \emph{Proc.
  Interspeech}, 2020, pp. 3426--3430.

\bibitem{hodariusing}
Z.~Hodari, O.~Watts, and S.~King, ``Using generative modelling to produce
  varied intonation for speech synthesis,'' in \emph{Proc. 10th ISCA Speech
  Synthesis Workshop}, 2019, pp. 239--244.

\bibitem{sun2020fully}
G.~Sun, Y.~Zhang, R.~J. Weiss, Y.~Cao, H.~Zen, and Y.~Wu, ``Fully-hierarchical
  fine-grained prosody modeling for interpretable speech synthesis,'' in
  \emph{Proc. ICASSP}, 2020, pp. 6264--6268.

\bibitem{tan2020fine}
D.~Tan and T.~Lee, ``Fine-grained style modeling, transfer and prediction in
  text-to-speech synthesis via phone-level content-style disentanglement,''
  \emph{arXiv preprint arXiv:2011.03943}, 2020.

\bibitem{wang2019vector}
X.~Wang, S.~Takaki, J.~Yamagishi, S.~King, and K.~Tokuda, ``A vector quantized
  variational autoencoder (vq-vae) autoregressive neural f0 model for
  statistical parametric speech synthesis,'' \emph{IEEE/ACM Trans. on Audio,
  Speech, and Language Processing}, vol.~28, pp. 157--170, 2019.

\bibitem{kenter2019chive}
T.~Kenter, V.~Wan, C.-A. Chan, R.~Clark, and J.~Vit, ``Chive: Varying prosody
  in speech synthesis with a linguistically driven dynamic hierarchical
  conditional variational network,'' in \emph{Proc. ICML}, 2019, pp.
  3331--3340.

\bibitem{zhang2021estimating}
G.~Zhang, S.~Qiu, Y.~Qin, and T.~Lee, ``Estimating mutual information in
  prosody representation for emotionalprosody transfer in speech synthesis,''
  in \emph{Proc. ISCSLP}, 2021, pp. 1--5.

\bibitem{hsu2019disentangling}
W.-N. Hsu, Y.~Zhang, R.~J. Weiss, Y.-A. Chung, Y.~Wang, Y.~Wu, and J.~Glass,
  ``Disentangling correlated speaker and noise for speech synthesis via data
  augmentation and adversarial factorization,'' in \emph{Proc. ICCASP}.\hskip
  1em plus 0.5em minus 0.4em\relax IEEE, 2019, pp. 5901--5905.

\bibitem{karlapati2020copycat}
S.~Karlapati, A.~Moinet, A.~Joly, V.~Klimkov, D.~Sáez-Trigueros, and
  T.~Drugman, ``{CopyCat: Many-to-Many Fine-Grained Prosody Transfer for Neural
  Text-to-Speech},'' in \emph{Proc. Interspeech}, 2020, pp. 4387--4391.

\bibitem{tishby2015deep}
N.~Tishby and N.~Zaslavsky, ``Deep learning and the information bottleneck
  principle,'' in \emph{Proc. Information Theory Workshop}, 2015, pp. 1--5.

\bibitem{alemi2016deep}
A.~Alemi, I.~Fischer, J.~Dillon, and K.~Murphy, ``Deep variational information
  bottleneck,'' in \emph{Proc. ICLR}, 2017.

\bibitem{tishby2000information}
N.~Tishby, F.~C. Pereira, and W.~Bialek, ``The information bottleneck method,''
  \emph{arXiv preprint physics/0004057}, 2000.

\bibitem{chechik2005information}
G.~Chechik, A.~Globerson, N.~Tishby, Y.~Weiss, and P.~Dayan, ``Information
  bottleneck for gaussian variables.'' \emph{Journal of machine learning
  research}, vol.~6, no.~1, 2005.

\bibitem{alemi2018fixing}
A.~Alemi, B.~Poole, I.~Fischer, J.~Dillon, R.~A. Saurous, and K.~Murphy,
  ``Fixing a broken elbo,'' in \emph{Proc. ICML}, 2018, pp. 159--168.

\bibitem{higgins2017learning}
I.~Higgins, L.~Matthey, A.~Pal, C.~Burgess, X.~Glorot, M.~Botvinick,
  S.~Mohamed, and A.~Lerchner, ``beta-vae: Learning basic visual concepts with
  a constrained variational framework.'' in \emph{Proc. ICLR}, 2017.

\bibitem{burgess2018understanding}
C.~P. Burgess, I.~Higgins, A.~Pal, L.~Matthey, N.~Watters, G.~Desjardins, and
  A.~Lerchner, ``Understanding disentangling in beta-vae.'' in \emph{Proc.
  ICLR}, 2018.

\bibitem{ren2019fastspeech}
Y.~Ren, Y.~Ruan, X.~Tan, T.~Qin, S.~Zhao, Z.~Zhao, and T.-Y. Liu, ``Fastspeech:
  Fast, robust and controllable text to speech,'' in \emph{Proc. NIPS}, 2019,
  pp. 3171--3180.

\bibitem{vaswani2017attention}
A.~Vaswani, N.~Shazeer, N.~Parmar, J.~Uszkoreit, L.~Jones, A.~N. Gomez,
  L.~Kaiser, and I.~Polosukhin, ``Attention is all you need,'' in \emph{Proc.
  NIPS}, 2017, pp. 3171--3180.

\bibitem{kingma2014auto}
D.~P. Kingma and M.~Welling, ``Auto-encoding variational bayes,'' \emph{stat},
  vol. 1050, p.~1, 2014.

\bibitem{fu2019cyclical}
H.~Fu, C.~Li, X.~Liu, J.~Gao, A.~Celikyilmaz, and L.~Carin, ``Cyclical
  annealing schedule: A simple approach to mitigating kl vanishing,'' in
  \emph{Proc. NAACL}, 2019, pp. 240--250.

\bibitem{van2017neural}
A.~Van Den~Oord, O.~Vinyals \emph{et~al.}, ``Neural discrete representation
  learning,'' in \emph{Proc. NIPS}, 2017, pp. 6306--6315.

\bibitem{baevski2019vq}
A.~Baevski, S.~Schneider, and M.~Auli, ``vq-wav2vec: Self-supervised learning
  of discrete speech representations,'' \emph{Proc. ICLR}, 2020.

\bibitem{ren2020fastspeech}
Y.~Ren, C.~Hu, T.~Qin, S.~Zhao, Z.~Zhao, and T.-Y. Liu, ``Fastspeech 2: Fast
  and high-quality end-to-end text-to-speech,'' \emph{Proc. ICLR}, 2021.

\bibitem{black1998festival}
P.~Taylor, A.~W. Black, and R.~Caley, ``The architecture of the festival speech
  synthesis system,'' in \emph{Proc. The Third ESCA/COCOSDA Workshop (ETRW) on
  Speech Synthesis}, 1998.

\bibitem{yamamoto2020parallel}
R.~Yamamoto, E.~Song, and J.-M. Kim, ``Parallel {WaveGAN}: {A} fast waveform
  generation model based on generative adversarial networks with
  multi-resolution spectrogram,'' in \emph{Proc. ICASSP}, 2020, pp. 6199--6203.

\bibitem{belghazi2018mutual}
M.~I. Belghazi, A.~Baratin, S.~Rajeshwar, S.~Ozair, Y.~Bengio, A.~Courville,
  and D.~Hjelm, ``Mutual information neural estimation,'' in \emph{Proc. ICML},
  2018, pp. 531--540.

\bibitem{devlin2019bert}
J.~Devlin, M.-W. Chang, K.~Lee, and K.~Toutanova, ``Bert: Pre-training of deep
  bidirectional transformers for language understanding,'' in \emph{Proc.
  NAACL}, 2019, pp. 4171--4186.

\bibitem{jawahar2019does}
G.~Jawahar, B.~Sagot, and D.~Seddah, ``What does bert learn about the structure
  of language?'' in \emph{Proc. ACL}, 2019, pp. 3651--3657.

\bibitem{ott2019fairseq}
M.~Ott, S.~Edunov, A.~Baevski, A.~Fan, S.~Gross, N.~Ng, D.~Grangier, and
  M.~Auli, ``fairseq: A fast, extensible toolkit for sequence modeling,'' in
  \emph{Proc. NAACL}, 2019, pp. 48--53.

\end{thebibliography}

\end{document}